\documentclass[a4paper,useAMS,usenatbib]{mn2e}

\usepackage{ifthen,psfig,graphicx}
\usepackage{epstopdf,aas_macros}
\usepackage{soul,color,amsmath}
\usepackage{rotating,longtable,caption}
\usepackage{dblfloatfix}
\usepackage{placeins}
\usepackage{multirow}
\def\msun{$M_{\odot}$} 
\def\lsun{$L_{\odot}$}

\def\mathnew{\mathsurround=0pt}
\def\simov#1#2{\lower 2.5pt\vbox{\baselineskip0pt \lineskip-.5pt
\ialign{$\mathnew#1\hfil##\hfil$\crcr#2\crcr\sim\crcr}}}

\newcommand{\MeV}{Me\kern-0.11em V}
\newcommand{\keV}{ke\kern-0.11em V}

\makeatletter
\newcommand\ion[2]{\mbox{#1$\;${\small\expandafter\@slowromancap\romannumeral #2@\relax}}}
\newcommand\iont[2]{{#1$\;${\small\expandafter\@slowromancap\romannumeral #2@\relax}}}
\makeatother
\makeatletter
\newcommand{\raisemath}[1]{\mathpalette{\raisem@th{#1}}}
\newcommand{\raisem@th}[3]{\raisebox{#1}{$#2#3$}}

\def\mathnew{\mathsurround=0pt}
\def\simov#1#2{\lower 2.5pt\vbox{\baselineskip0pt \lineskip-.5pt
\ialign{$\mathnew#1\hfil##\hfil$\crcr#2\crcr\sim\crcr}}}
\def\lesssim{\mathrel{\mathpalette\simov <}}

\makeatother

\title[Large Decay of X-ray Flux in  2XMM J123103.2+110648: Evidence for a
Tidal Disruption Event]{Large Decay of X-ray Flux in  2XMM
  J123103.2+110648: Evidence for a Tidal Disruption Event }
\author[D. Lin et al.]{Dacheng Lin$^{1}$\thanks{E-mail: dacheng.lin@unh.edu}, Olivier Godet$^{2,3}$,  Luis C. Ho$^{4,5}$,  Didier Barret$^{2,3}$,  Natalie A. Webb$^{2,3}$, \newauthor Jimmy A. Irwin$^{6}$
  \smallskip\\
  $^{1}$Space Science Center, University of New Hampshire, Durham, NH 03824, USA\\
  $^{2}$CNRS, IRAP, 9 avenue du Colonel Roche, BP 44346, F-31028 Toulouse Cedex 4, France\\
  $^{3}$Universit\'{e} de Toulouse, UPS-OMP, IRAP, Toulouse, France\\
  $^{4}$Kavli Institute for Astronomy and Astrophysics, Peking University, Beijing 100871, China\\
  $^{5}$Department of Astronomy, School of Physics, Peking University, Beijing 100871, China\\
  $^{6}$University of Alabama, Department of Physics and Astronomy, Tuscaloosa AL 35487, USA
}
\begin{document}

\date{In original form 2016 December 10}

\pagerange{\pageref{firstpage}--\pageref{lastpage}} \pubyear{2017}

\maketitle

\label{firstpage}

\begin{abstract}
  The X-ray source 2XMM J123103.2+110648 was previously found to show
  pure thermal X-ray spectra and a $\sim$3.8 hr periodicity in three
  \textit{XMM-Newton} X-ray observations in 2003--2005, and the
  optical spectrum of the host galaxy suggested it as a type 2 active
  galactic nucleus candidate. We have obtained new X-ray observations
  of the source, with \textit{Swift} and \textit{Chandra} in
  2013--2016, in order to shed new light on its nature based on its
  long-term evolution property. We found that the source could be in
  an X-ray outburst, with the X-ray flux decreasing by an order of
  magnitude in the \textit{Swift} and \textit{Chandra} observations,
  compared with the \textit{XMM-Newton} observations ten years
  ago. There seemed to be significant spectral softening associated
  with the drop of X-ray flux (disk temperature $kT\sim0.16$--0.2 keV
  in \textit{XMM-Newton} observations versus $kT\sim0.09\pm0.02$ keV
  in the \textit{Chandra} observation). Therefore the \textit{Swift}
  and \textit{Chandra} follow-up observations support our previous
  suggestion that the source could be a tidal disruption event (TDE),
  though it seems to evolve slower than most of the other TDE
  candidates. The apparent long duration of this event could be due to
  the presence of a long super-Eddington accretion phase and/or slow
  circularization.

\end{abstract}

\begin{keywords}
accretion, accretion disks --- black hole physics --- X-rays: galaxies -- galaxies: individual: 2XMM J123103.2+110648
\end{keywords}

\section{Introduction}

Active galactic nuclei (AGN) typically show hard X-ray spectra
  (photon index $\sim$2.0 when 2--10 keV spectra are fitted with an
  absorbed power-law), in addition to a possible soft excess below
  around 1.0 keV
  \citep{tupo1989,napo1994,wiel1987,coseza1992,liweba2012,gido2004}.
Very few galactic nuclei exhibit pure thermal X-ray spectra with the
hard X-ray component either extremely weak or completely absent. Most
of these outliers are observed in the candidate tidal disruption
events (TDEs), in which stars are tidally disrupted and subsequently
accreted by massive black holes (BHs) at the center of galaxies
\citep[for recent reviews, refer to][]{ko2012,ko2015}. About 30 such
TDEs have been found, with host galaxies showing no sign or weak sign
of persistent nuclear activity in the optical spectra. However, pure
thermal X-ray spectra were also detected in several galactic nuclei
with clear narrow emission lines in optical, suggesting possible
persistent nuclear activity. Good examples of such objects include
GSN 069 \citep{misaro2013}, 2XMM J123103.2+110648
\citep[XJ1231+1106
  hereafter,][]{liweba2012,liirgo2013,liweba2014,tekaaw2012,hokite2012},
and IC 3599 \citep{grbema1995,grkosa2015,camaco2015}.

GSN 069 was discovered in 2010 and was found to be in an outburst,
with the X-ray flux a factor of $>$240 higher than ROSAT observations
in the early 1990s. The outburst seems to be semi-persistent, with the
flux remaining fairly steady since it was discovered
\citep{misaro2013}. The X-ray spectra were supersoft, consistent with
sub-Eddington thermal disk emission from a BH of mass $\sim$$10^6$
\msun\ \citep{misaro2013}. The optical spectrum exhibited no broad
emission lines but only narrow ones suggesting a low-luminosity
Seyfert 2 galaxy.

XJ1231+1106 was serendipitously detected by \emph{XMM-Newton} in two
epochs separated by 2.4 yr, with the luminosity slightly lower in the
first epoch than in the second one. In the second epoch there were two
observations four days apart, with a $\sim$3.8 hr quasi-periodic
oscillation significantly ($\sim$5$\sigma$) detected in both of them
\citep{liirgo2013}. The optical spectrum of its host galaxy
SDSS J123103.24+110648.6 (GJ1231+1106 hereafter) also exhibited no
broad emission lines but narrow ones consistent with a low-luminosity
AGN \citep{hokite2012}. The width of the narrow lines is so small
(velocity dispersion $\sigma=34$ km s$^{-1}$ for [\ion{O}{3}]
$\lambda$5007) that \citet{hokite2012} inferred the BH mass to be only
$\sim$$10^5$ \msun.

IC 3599 has at least two outbursts since it was discovered in
  1990 \citep{grbema1995,grkosa2015,camaco2015}. It had ultrasoft
  X-ray spectra in the peak of the outbursts, though in the low state
  it behaved as a typical AGN. The repeated outbursts were
  explained as recurrent partial disruption of a star by the central
  SMBH \citep{camaco2015} or as AGN flares caused by a disk
  instability \citep{grkosa2015}.

In this paper we continue to study XJ1231+1106. The source evolution
on timescales longer than the \emph{XMM-Newton} observations was
unclear. Although it was not detected in the \textit{ROSAT} All-Sky Survey, the
non-detection could be just because of the low sensitivity of the
survey, which had a detection limit of
  5$\times$10$^{-13}$ erg s$^{-1}$ cm$^{-2}$ in 0.1--2.4 keV \citep{voasbo1999}, five
  times higher than the fluxes of the source in the \emph{XMM-Newton}
  observations.  We have obtained \emph{Swift} and \emph{Chandra}
follow-up observations of the source and found a significant decrease
in the X-ray flux. We report this finding in this Letter. Here we also
fit the optical spectrum of the host galaxy GJ1231+1106 taken by the
SDSS, aiming to reveal more properties of the environment of
XJ1231+1106.  In Section~\ref{sec:reduction}, we describe the data
analysis. In Section~\ref{sec:res}, we present the results. The
discussion of the source nature and our conclusions are given in
Section~\ref{sec:conclusion}.

\begin{figure*}
\begin{center}
\includegraphics[width=5.2in]{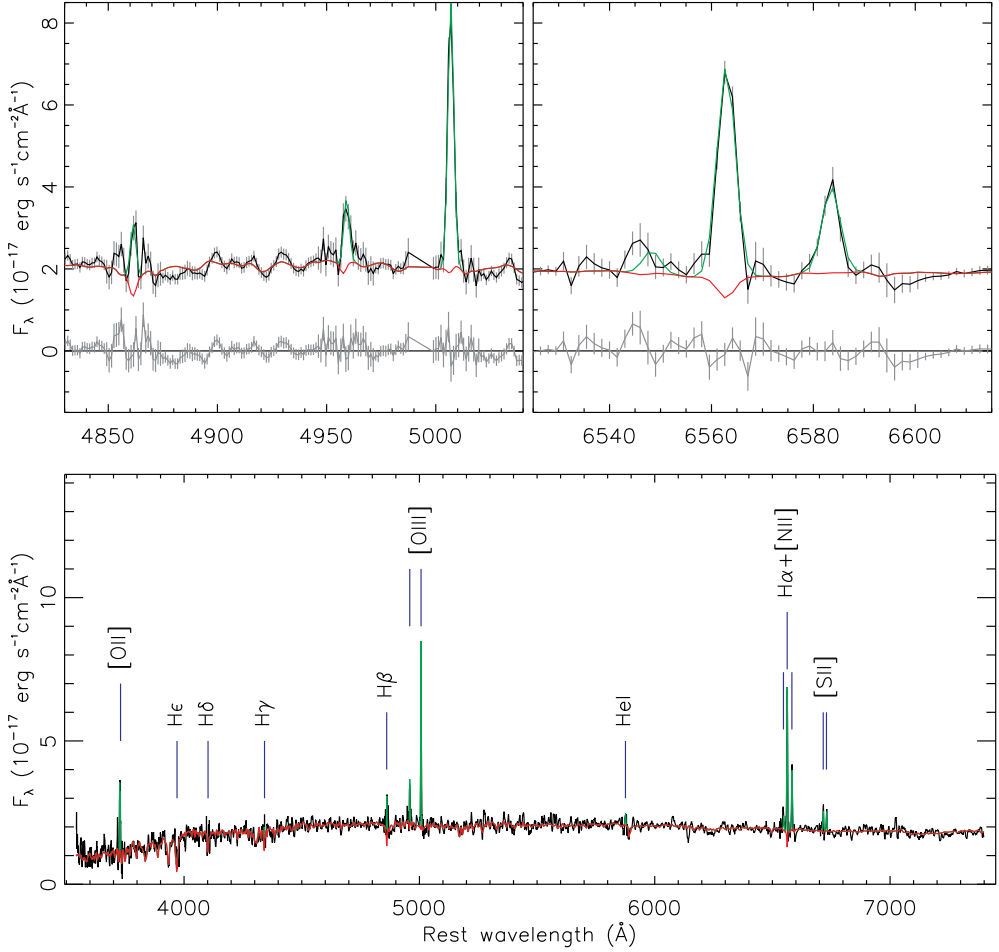}
\end{center}
\vskip -0.2in
\caption{The SDSS optical spectrum of the candidate host galaxy of XJ1231+1106 taken on 2012 March 3, showing only narrow emission lines. The upper two panels zoom into the H$\beta$-[\ion{O}{3}] complex and the H$\alpha$-[\ion{N}{2}] region, with the fit residuals. The pPXF fit is shown as a solid green line, while the star component is shown as a red line. The data points outside the emission line regions have been smoothed with a box function of width 5 for clarity. \label{fig:sdsssp}}
\end{figure*}

\begin{figure*}
\begin{center}
\includegraphics[width=5.2in]{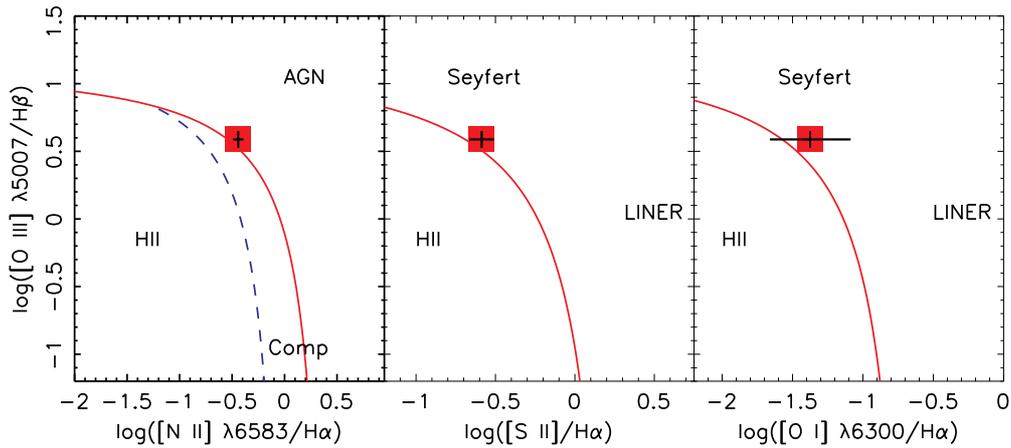}
\end{center}
\vskip -0.2in
\caption{XJ1231+1106 on the BPT diagrams, indicating it as an AGN possibly with some star-forming activity. The
  dashed and solid lines are used to separate galaxies into
  HII-region-like, AGN, and composite
  types \citep{kegrka2006} \label{fig:bpt}}
\end{figure*}

\begin{table}
\centering \caption{X-ray observations and spectral fit results$^{a}$.} \label{tbl:obslog}
\begin{tabular}{lllrl}
\hline
Obs.\,ID & Start date & $kT_{\rm MCD}$  & $F_{\rm X}$$^{b}$ & $F_{\rm bol}$$^{c}$ \\
&  & (keV) &  \multicolumn{2} {c} {($10^{-13}$ erg s$^{-1}$ cm$^{-2}$)}  \\
\hline
\multicolumn{2} {l} {\textit{XMM-Newton}:}\\
0145800101  & 2003--07--13 &  $ 0.16^{+ 0.01}_{-0.01}$ & $ 0.65^{+ 0.11}_{-0.06}$  & $ 1.48^{+ 0.33}_{-0.20}$   \\
0306630101 & 2005--12--13 &   $ 0.20^{+ 0.01}_{-0.01}$ & $ 1.38^{+ 0.18}_{-0.10}$  & $ 2.55^{+ 0.45}_{-0.24}$   \\
0306630201  & 2005--12--17 &  $ 0.18^{+ 0.01}_{-0.01}$ & $ 1.04^{+ 0.15}_{-0.08}$  & $ 2.17^{+ 0.44}_{-0.23}$ \\
\hline
\textit{Swift}:\\
00032732001 & 2013-03-08& \multirow{11}{*}{$ 0.15^{+ 0.07}_{-0.05}$} & \multirow{11}{*}{$ 0.16^{+ 0.09}_{-0.07}$}  & \multirow{11}{*}{$ 0.39^{+ 0.44}_{-0.20}$} \\
00032732002 & 2013-06-21\\
00032732003 & 2013-11-21\\
00032732004 & 2014-01-04\\
00032732005 & 2014-01-08\\
00032732006 & 2014-02-13\\
00032732007 & 2014-03-29\\
00032732008 & 2014-05-08\\
00032732009 & 2014-05-13\\
00032732010 & 2014-06-17\\
00032732011 & 2014-07-27\\
\hline
\textit{Chandra}:\\
17129 & 2016-02-10 & $ 0.09^{+ 0.03}_{-0.02}$  & $ 0.09^{+ 0.05}_{-0.04}$  & $ 0.52^{+ 0.59}_{-0.29}$\\
\hline
\end{tabular}
\begin{list}{}{}
\item[$^{a}$] The fits used the MCD model. All uncertainties are at the 90\% confidence level. The fits to \textit{XMM-Newton} observations are from \citet{liirgo2013} and we refer to it for more details.
\item[$^{b}$] Unabsorbed 0.34--11 keV  (source rest-frame) flux.
\item[$^{c}$] Unabsorbed bolometric flux (based on the MCD component).
\end{list}
\end{table}

\begin{figure*}
\begin{center}
\includegraphics[width=5.2in]{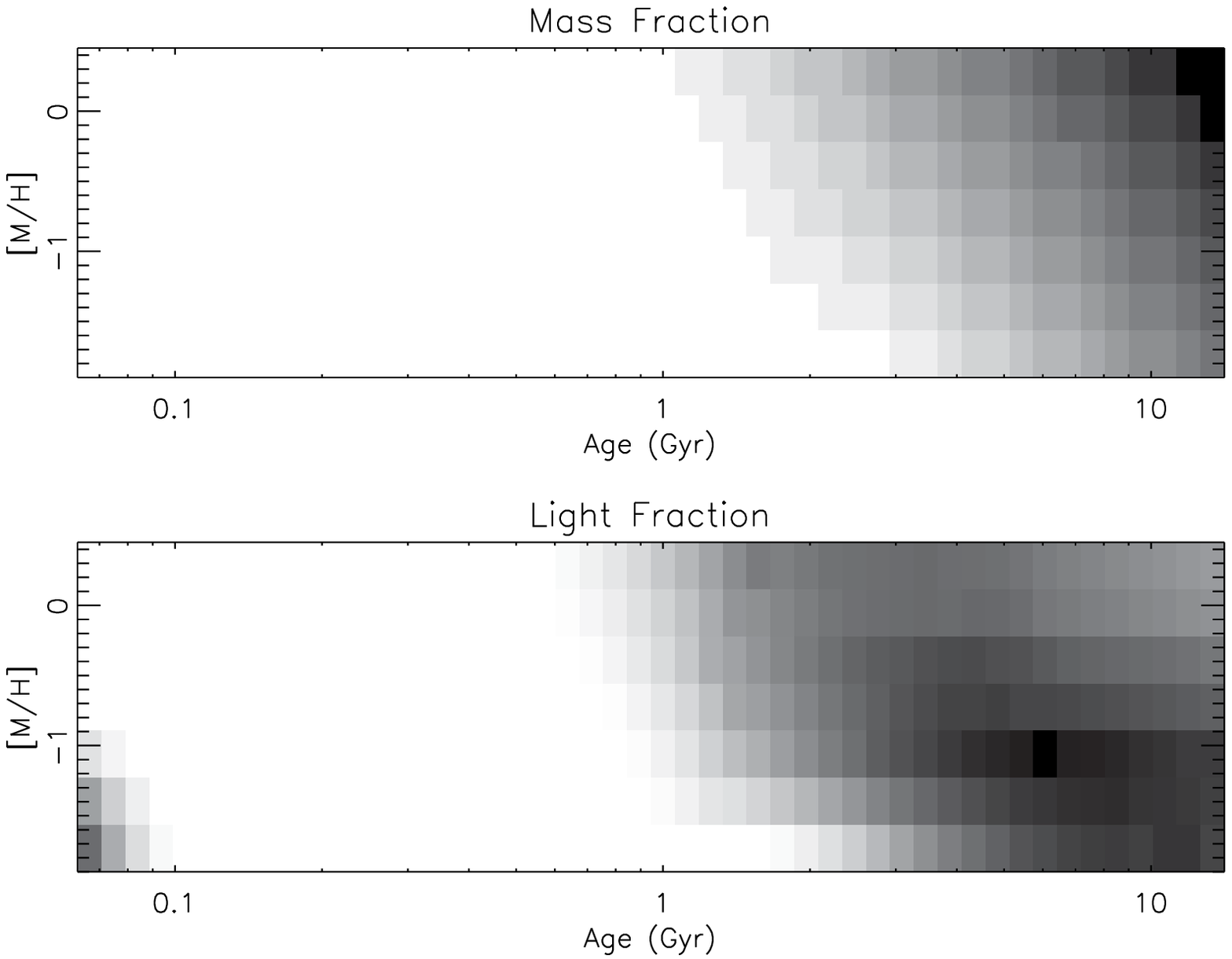}
\end{center}
\vskip -0.2in
\caption{Relative mass and light fractions of different stellar
  populations in the host galaxy of XJ1231+1106 with respect to
  metallicity and age, with darker shading indicating a larger mass
  fraction in the best-fitting model. There seems to be a young
  ($<90$ Myr) population, based on the light fraction plot. The light was integrated over 3540 \AA\ and
  7410 \AA.
  %The inset plot zooms in on the H$\alpha$ region. The red dashed line is the expected broad H$\alpha$ component (plus the continuum) assuming a BH with $M_{\rm BH}=10^5$ \msun\ and a FWHM of 500 km~s$^{-1}$, while the green dotted line assumes a BH with $M_{\rm BH}=10^6$ \msun\ and a FWHM of 1000 km~s$^{-1}$ (see the text for details).
  \label{fig:masslightdis}}
\end{figure*}

\begin{figure}
\begin{center}
\includegraphics[width=3.4in]{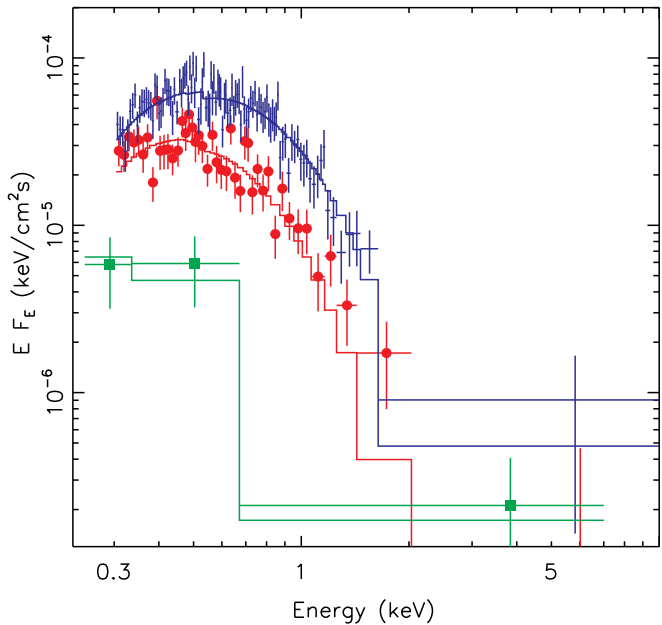}
\end{center}
\vskip -0.2in
\caption{The example unfolded spectra of XJ1231+1106 fitted with the
  MCD model. From the top to the bottom are the second
  \textit{XMM-Newton} observation (blue), the first
  \textit{XMM-Newton} observation (July 13rd 2003, red filled
  circles), and the \textit{Chandra} observation (green filled
  squares). For clarity, we show the pn data only for the
  \textit{XMM-Newton} observations and have rebinned all spectra to be
  above $2\sigma$ per bin.
\label{fig:spectralfit}}
\end{figure}

\begin{figure}
\begin{center}
\includegraphics[width=3.4in]{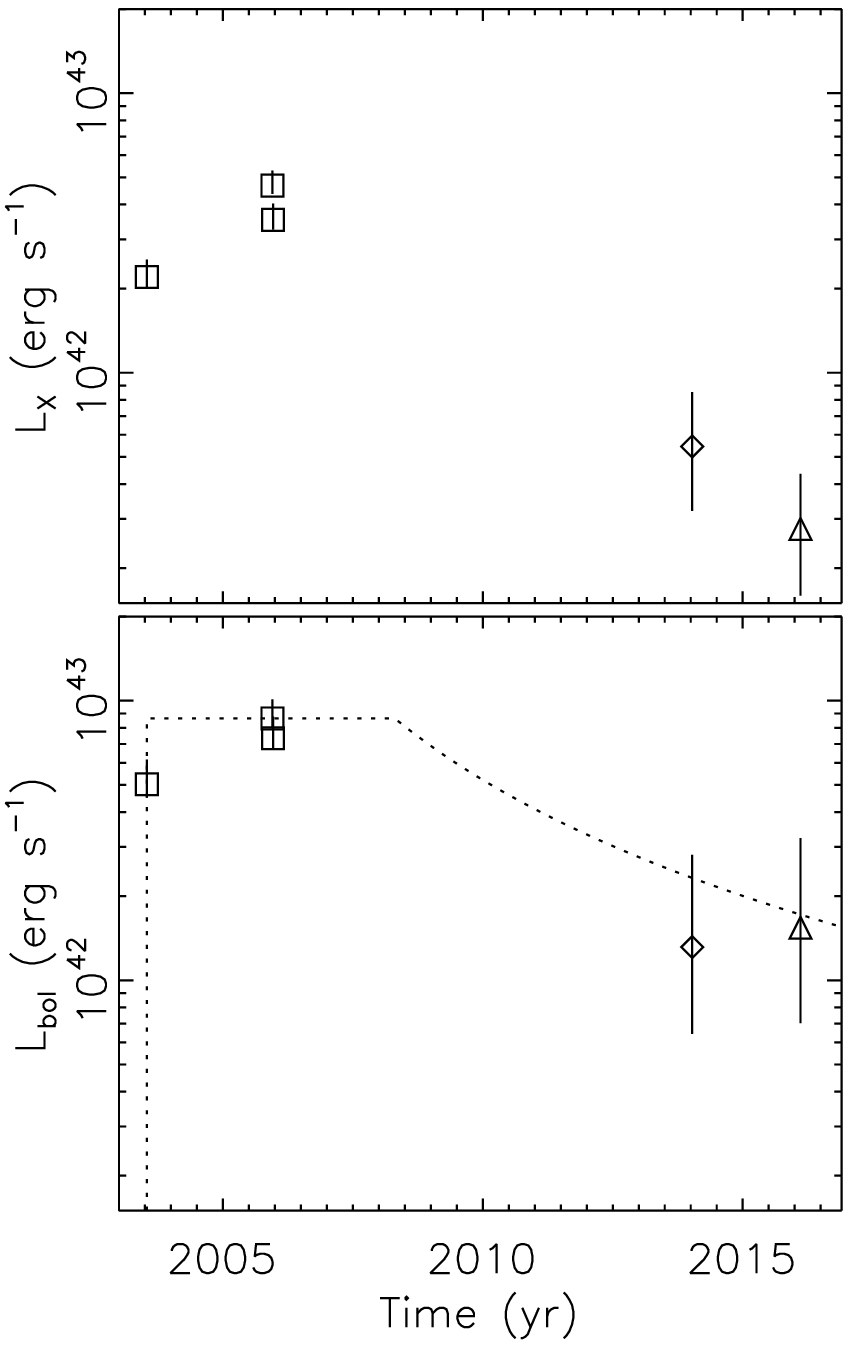}
\end{center}
\vskip -0.2in
\caption{The long-term luminosity curve of XJ1231+1106 from the
  \textit{XMM-Newton} (squares), \textit{Swift} (diamond),
  \textit{Chandra} (triangle) observations. The upper panel is for the
  0.34--11 keV (source rest-frame) unabsorbed luminosity, and the
  bottom panel is for the bolometric luminosity based on the MCD
  model. The dashed line plots a TDE model of prompt accretion (see
  Section~\ref{sec:conclusion} for more details).
  \label{fig:ltlumlc}}
\end{figure}

\section{DATA ANALYSIS}
\label{sec:reduction}
\subsection{\textit{Swift} Observations}
At our request, \textit{Swift} \citep{gechgi2004} carried out 11
observations of XJ1231+1106 between 2013 March and 2014 July
(Table~\ref{tbl:obslog}). The first two observations had been studied
in \citet{liirgo2013}, and here we analyze all observations with
FTOOLS 6.19 and updated calibration files (released on 2016 November
1). The X-ray telescope \citep[XRT;][]{buhino2005} was operated in
Photon Counting mode.  We reprocessed the event files with the task
\texttt{xrtpipeline} (version 0.13.2). The source was not clearly
detected in any observation. In order to probe a deeper sensitivity,
we created a co-added spectrum from all observations (total exposure
time 51 ks) using a source region of radius 20$\arcsec$ and a
background region of radius 2$\arcmin$. The UV-Optical Telescope
\citep[UVOT;][]{rokema2005} used the UVW1 filter (8.3 ks) in the first
observation and the UVW2 filter in the other observations (total
exposure 41.1 ks). The UV magnitudes and fluxes were measured with the
task \textit{uvotsource} with a source region of radius 5$\arcsec$ and
a background region of radius 25$\arcsec$. This is done both for
individual observations and for the merged observation (the UVW2
filter).

\subsection{\textit{Chandra} Observation}
We had a follow-up observation of XJ1231+1106 with \textit{Chandra} on
2016 February 10 (Table~\ref{tbl:obslog}). It used the imaging array
of the AXAF CCD Imaging Spectrometer \citep[ACIS; ][]{bapiba1998},
with the aim point at the back-illuminated chip S3, given that the
source had been ultrasoft.  We reprocessed the data with the script
\textit{chandra\_repro} in the \textit{Chandra} Interactive Analysis
of Observations (CIAO, version 4.8) package and applied the latest
calibration (CALDB 4.7.2). The exposure time of the observation is
39.5 ks, without background flares. We extracted the source and
background spectra and created the corresponding response matrices for
all observations using the script \textit{specextract}. The radius of
the source region used was 1\farcs6, corresponding to PSF enclosing
fractions of 95\% at 1.0 keV, and the radius of the background region
was $30\arcsec$.

\subsection{The SDSS Spectroscopic Observation}
The SDSS took a spectrum of the host galaxy of XJ1231+1106 on 2012
January 23. We fitted the spectrum with multi-component models comprised
of single-population synthetic spectra, using Penalized Pixel Fitting
({\tt pPXF}) software \citep{caem2004} and \cite{vasafa2010} synthetic
spectra spanning a grid of 48 ages between 0.06 to 14 Gyr and 7
metallicities [M/H]=\{$-2.32$, $-1.71$, $-1.31$, $-0.71$, $-0.40$,
$0.00$, $+0.22$\}. The spectrum also exhibits narrow emission lines,
which were fitted with Gaussian functions. No additive or
multiplicative polynomials were included in the fit. The spectrum was
corrected for the Galactic dust reddening \citep{scfida1998} of
$\mathrm{E(B-V)}_\mathrm{G}=0.03$ mag before the fit. The intrinsic
reddening was left as a free fitting parameter and was inferred to be
negligible.

\section{Results}
\label{sec:res}
\subsection{The Host Galaxy}

The SDSS spectrum of the host galaxy GJ1231+1106 is shown in
Figure~\ref{fig:sdsssp}. The pPXF fit of the star component is shown
as a red line. The light-weighted age is 5.9 Gyr, and the
mass-weighted age is 7.8 Gyr. The total stellar mass is
$\sim$$5.5\times10^{9}$ \msun, and the total luminosity within the
fitting band (source rest-frame 3540-7410 \AA) is
$\sim$$7.8\times10^{8}$ \lsun, indicating a dwarf host galaxy (mass
comparable to the Large Magellanic Cloud). Based on the relation
between the BH mass and the total galaxy stellar mass from
\citet{revo2015}, we estimated the BH mass to be $1.5\times10^6$
\msun\ (an intrinsic scatter of 0.55 dex). The mass and light
distributions of the stellar populations are shown in
Figure~\ref{fig:masslightdis}. There seems to be a young population
of age $<$90 Myr, suggesting some level of star-forming activity.

The pPXF fit inferred the intrinsic width of the emission lines to be
$\sigma=43\pm4$ km s$^{-1}$. This value is below the SDSS instrumental
resolution ($70$ km s$^{-1}$) and thus should be taken with caution,
though it is consistent with that obtained by \citet{hokite2012},
using a Magellan spectrum with a much higher resolution. The intrinsic
reddening of the emission lines is $\mathrm{E(B-V)}_\mathrm{i}=0.29$
mag, assuming an intrinsic H$\alpha$/H$\beta$ ratio of 3.1. Accounting
for this small intrinsic reddening, the line ratios put the source in
the Seyfert region on the BPT diagrams
\citep[Figure~\ref{fig:bpt},][]{baphte1981,veos1987}, but close to the
boundary between the HII and Seyfert regions. This indicates some
level of star-forming activity, which is consistent with the fit of the star
component, and/or the low metallicity of the host
\citep[e.g.,][]{lugrba2012}.

There is a faint UV source from the \textit{Swift} observation at the
position of GJ1231+1106. We obtained the W1 magnitude of $22.6\pm0.2$
AB mag (flux $1.4\pm0.2\times10^{-17}$ erg s$^{-1}$ cm$^{-2}$
\AA$^{-1}$) in the first \textit{Swift} observation. The mean W2
magnitude is $23.2\pm0.1$ AB mag (flux $1.3\pm0.1\times10^{-17}$ erg
s$^{-1}$ cm$^{-2}$ \AA$^{-1}$) from the merged observation. The ten
individual observations gave consistent W2 magnitudes (mostly within
$1\sigma$ but none above $3\sigma$). Therefore we did not detect
significant variability in the W2 filter. Based on the W1 and W2 photometry, the UV source seems blue and can be explained with the
presence of some star-forming activity.

\subsection{X-ray Follow-up}
In the \textit{Chandra} follow-up observation in 2016, we obtained 13
counts within 0.24--7 keV, with 0.7 background counts, at the position
of XJ1231+1106. The net source count rate is $3.1\pm0.9\times10^{-4}$
counts s$^{-1}$ ($1\sigma$ uncertainty), a factor of 26 lower than
expected from the second \textit{XMM-Newton} spectrum (obtained on
2005 December 13). The \textit{Chandra} spectrum seems very soft, with
11 counts below 0.7 keV (background is expected to be negligible). We
rebinned the source spectrum to have at least one count per bin and
carried out the fit with the multi-color disk (MCD) model and adopting
the C statistic in XSPEC. The Galactic absorption was fixed at
$N_\mathrm{H}=2.3\times10^{20}$ cm$^{-2}$ \citep{kabuha2005}, and the
intrinsic absorption was fixed at $N_{\rm H,i}=6\times10^{19}$ cm$^{-2}$
as obtained in \citet{liirgo2013}. The fit is shown in
Figure~\ref{fig:spectralfit} and given in Table~\ref{tbl:obslog}. We inferred a source rest-frame disk
temperature of $kT_\mathrm{MCD}=0.09\pm0.02$ keV (90\%
uncertainty). Given the low number of counts of the spectrum, we
checked whether the fit was valid by simulating 1000 spectra of the
same counts based on the best-fitting MCD model and then fitting with
the same MCD model. We found that the best-fitting $kT_\mathrm{MCD}$
from the simulated spectra had the median and the scatter fully
consistent with the best-fitting value and uncertainty of
$kT_\mathrm{MCD}$ obtained above from the fit to the observed
spectrum. Therefore, the source X-ray spectrum is not only fainter but
also softer, at the confidence level of $\sim5\sigma$, compared with
the last \textit{XMM-Newton} observation
\citep[$kT_\mathrm{MCD}=0.18\pm0.01$ keV,][]{liirgo2013}. We note that
the intrinsic absorption used to fit the \textit{Chandra}
spectrum was so low that assuming zero absorption would not change the
inferred disk temperature, while assuming stronger absorption would
infer a lower disk temperature. Therefore our conclusion of the spectral
softening is not subject to the intrinsic absorption assumed.

The source is also weakly detected in the combined \textit{Swift}
observations in 2013--2014. There are 28 counts within 0.3--10 keV,
with 13 background counts expected (so 15 net source counts). The net
source count rate is $2.9\pm1.0\times10^{-4}$ counts s$^{-1}$
($1\sigma$ uncertainty), a factor of 11 lower than expected from the
second \textit{XMM-Newton} spectrum. The source spectrum was also
soft, with 12.5 net source counts below 0.7 keV. We also carried out a
MCD fit to this \textit{Swift} spectrum and obtained
$kT_\mathrm{MCD}=0.15\pm0.05$ keV (Table~\ref{tbl:obslog}). The large
uncertainty makes it hard to conclude whether the source spectrum was
softer than \textit{XMM-Newton} observations, though it was
significantly fainter.

The long-term luminosity evolution of the source based on the MCD fits
is shown in Figure~\ref{fig:ltlumlc}. The source could be in an X-ray
outburst, with the peak X-ray luminosity of $\sim4\times10^{42}$ erg
s$^{-1}$, reached in the second and third \textit{XMM-Newton}
observations. The X-ray luminosity had decreased to
$\sim$$3\times10^{41}$ erg s$^{-1}$ in the \textit{Chandra}
observation. The bolometric luminosity was $\sim10^{43}$ erg s$^{-1}$
in these observations and is fairly independent of the spectral model
(an MCD or optically thick low-temperature corona). The bolometric
luminosity had decreased to $\sim$~$2\times10^{42}$ erg s$^{-1}$ in
the \textit{Chandra} observation .

\section{Discussion and Conclusions}
\label{sec:conclusion}
\subsection{The AGN Explanation}

The main result of our follow-up observation of XJ1231+1106 is the
detection of the significant drop and softening of its X-ray emission,
suggesting that the source is probably in the decay of an X-ray
outburst that has last for $>$13 years. Because the narrow emission
lines in the host galaxy spectrum signal the presence of a
low-luminosity type-2 AGN, in principle we cannot completely rule out
that the X-ray outburst is purely due to AGN activity. However, pure
thermal X-ray spectra are atypical for AGNs
\citep[e.g.,][]{liweba2012}. The large variability of the source is
uncommon for AGNs too \citep[only 1.5\% of AGNs vary in X-rays by a
  factor of $>$10,][]{liweba2012}.

One possible explanation for XJ1231+1106 is an AGN flare
  caused by a disk instability, as proposed for NGC 3599
  \citep{samoko2015} and IC 3599 \citep{grkosa2015}. The disk
  instability can cause large variability on timescales of years for
  AGNs. However, we note that this mechanism is still poorly
  understood and that the observational evidence is still vague. For
  NGC 3599 and IC 3599, alternative explanations, especially TDEs, are
  still possible \citep{samoko2015,camaco2015}.

Another possible explanation for the X-ray outburst of XJ1231+1106
is an AGN just entering a luminous thermal state, a common
spectral state for BH X-ray binaries, in which the X-ray spectra were
dominated by emission from a standard thermal thin disk. This
explanation was proposed to explain the super-soft semi-persistent
X-ray outburst of GSN 069 \citep{misaro2013}. However, such a state
would last $>10^4$ years in AGNs if the analogy between BH X-ray
binary transients and AGN holds, and this is in conflict with our
detection of the large variability of the source flux within 10 years.

\subsection{The TDE Explanation}

Pure thermal X-ray spectra are commonly observed in sources suspected
to be TDEs.  Therefore \citet{liirgo2013} discussed XJ1231+1106 as a
possible TDE. This is further supported by our new
detection of the outburst-like large variability of the source. One
main difference between XJ1231+1106 and other TDE candidates is its
slow evolution. The X-ray flux of other TDE candidates typically
decreased by an order of magnitude in one year right after they were
discovered \citep[e.g.,][]{kohasc2004,mauler2013}, while it took 10
years for the X-ray flux of XJ1231+1106 to decrease by one order of
magnitude. The first \textit{XMM-Newton} observation was even fainter
(by a factor of 2) than the other two \textit{XMM-Newton} observations
2.4 years later. The X-ray spectral softening observed in XJ1231+1106
with the decrease in the X-ray flux has been observed in some TDE
candidates \citep{licagr2011,sarees2012}, but not in all cases. In the
well monitored TDE candidate ASASSN-14li, steady blackbody
temperatures in the X-ray spectra were observed despite the drop of
the X-ray flux nearly by an order of magnitude \citep{mikami2015}. It
is not clear what caused the above difference.

The slow decay and spectral softening of the X-ray emission makes
XJ1231+1106 somewhat similar to a long-lived TDE candidate
3XMM~J150052.0+015452, which showed little decay of the X-ray flux
over 10 years after it went into an X-ray outburst and displayed
dramatic spectral softening, from quasi-soft ($kT\sim0.3$ keV) to
super-soft ($kT\sim0.13$ keV) X-ray spectra \citep{liguko2017}. One
explanation for the slow decay of that source is the combination of
slow circularization and super-Eddington accretion effects. Slow
circularization occurs when the fallback of mass is faster than the
accretion and have been predicted in many recent studies
\citep{ko1994,gura2015,pisvkr2015,shkrch2015,hastlo2016}. Such events
would evolve slower than those with prompt accretion.  TDEs could have
accretion rates above the Eddington limit initially. In this phase,
significant super-Eddington effects of photon trapping and outflows in
the inner disk region are expected
\citep{ohmi2007,krpi2012,kimu2016}. These effects are stronger at
higher accretion rates, resulting in a disk luminosity sustained at
around the Eddington limit. Therefore the decay of the flux would
appear slow in the super-Eddington accretion phase. The identification
of a long super-Eddington accretion phase in 3XMM~J150052.0+015452 was
strongly supported by the generally quasi-soft X-ray spectra, whose
characteristic temperatures are too high to be explained by the
standard thermal thin disk below the Eddington limit, but are
consistent with Comptonized emission from a low-temperature optical
thick corona. Similar spectra are commonly seen in ultraluminous X-ray
sources (ULXs) \citep{glrodo2009,liirwe2013,mimima2013}, most of which
are believed to be super-Eddington accreting stellar-mass BHs, except
that 3XMM~J150052.0+015452 had orders of magnitude higher
luminosities.

XJ1231+1106 was most likely in the thermal state, instead of a
super-Eddington accretion state, in the \textit{Swift} and
\textit{Chandra} observations, given the much lower luminosity and
softer spectra in these observations than in the \textit{XMM-Newton}
ones. The spectral state identification for the \textit{XMM-Newton}
observations is more subtle. \citet{liirgo2013} found that the
\textit{XMM-Newton} X-ray spectra can be described with either pure
thermal disk emission (i.e., the standard thermal state) or optically
thick low-temperature Comptonization (characteristic of a
super-Eddington accretion state). The former model requires a
sub-Eddington luminosity. Due to the relatively high disk
temperatures, this model required the BH to have mass $\sim$10$^5$
\msun\ and some spin. If the source was instead in a super-Eddington
accretion state in the \textit{XMM-Newton} observations, the BH should
have mass below $\sim10^5$ \msun\ (but above $10^4$ \msun, so that it
can be in the standard thermal state in the \textit{Swift} and
\textit{Chandra} observations). In either case, the BH mass should be
small, agreeing with the estimate by \citet{hokite2012} and a little
lower than our estimate above based on the stellar mass (the
difference is not significant due to large scatter of the relation
used).

In Figure~\ref{fig:ltlumlc}, we plot a model of the luminosity
evolution assuming a full disruption of a solar-type star by a SMBH of
mass $10^5$ \msun, with prompt accretion (i.e., the mass accretion
rate is equal to the mass fallback rate). In this case, the rising
time is expected to be $\lesssim$1 month \citep{ul1999,gura2015}. The
first \textit{XMM-Newton} observation should be in the rising phase
because the source had a lower luminosity in this observation than in
the two observations 2.4 years later. The super-Eddington accretion
phase can last for 4.8 years \citep{ul1999}. Therefore the source
luminosity was assumed to be constant in the initial 4.8 years, at the
Eddington limit \citep{krpi2012}, which was adopted to be that seen in
the second \textit{XMM-Newton} observation. After the super-Eddington
accretion phase, the luminosity followed the standard evolution of the
mass accretion rate as $(t-t_{\rm d})^{-5/3}$ \citep{re1988,ph1989},
where $t_{\rm d}$ is the disruption time, assumed to be
  one month before the first \textit{XMM-Newton} observation.

This simple model of a TDE of prompt accretion seems to describe the
data well. However, it would be highly coincident that the first
\textit{XMM-Newton} observation caught the fast rise. Alternatively,
it could be a TDE of slow circularization and thus of a relatively
long rising phase \citep{gura2015}. Tidal stripping of an
  evolved star could also result in a relatively slow event
  \citep{magura2012}. However, TDEs of evolved stars should be very
  rare for a black hole of mass $10^5$ \msun, accounting for only
  $\sim$3\% of the total TDEs \citep{ko2016}. Besides, in such events,
  the partial disruption is more likely, resulting in low luminosities
  \citep{maragr2013}, while our source had peak luminosity around the
  Eddington limit. 

As shown in the BPT diagrams, the narrow emission lines in the SDSS
spectrum of GJ1231+1106 could in some part be due to persistent
nuclear activity. Then we expect the presence of a hard X-ray
component if it is a normal AGN. The strength of this hard X-ray
component can be estimated based on its normal correlation with the
[\ion{O}{3}] $\lambda$5007 emission line \citep{labima2009}. The
[\ion{O}{3}] $\lambda$5007 emission line absorbed luminosity is
$8.7\pm0.4\times10^{39}$ erg s$^{-1}$. The observed H$\alpha$/H$\beta$
ratio is $4.3\pm0.6$, and assuming an intrinsic value of the ratio to
be 3.1 would imply an intrinsic reddening of
$\mathrm{E(B-V)}_\mathrm{i}=0.30\pm0.14$ mag \citep[Galactic reddening
  $\mathrm{E(B-V)}_\mathrm{G}=0.03$ mag,][]{scfida1998}. Correcting
for this reddening, the [\ion{O}{3}] $\lambda$5007 luminosity is
$2.5\pm1.1\times10^{40}$ erg s$^{-1}$. Using the [\ion{O}{3}]
$\lambda$5007 and 2--10 keV luminosity relation in \citet{labima2009},
whose dispersion is 0.63 dex, we estimated the persistent 2--10 keV
luminosity to be $3.1\pm2.4\times10^{41}$ erg s$^{-1}$ (the $1\sigma$
uncertainty has included the dispersion of the relation
used). Assuming an absorbed power-law of photon index 2.0 and Galactic
absorption, we expect to collect $12\pm9$ counts in 2--7 keV and
$45\pm35$ counts in 0.7--7 keV in the \textit{Chandra} observation but
only 1.5 net counts were detected in 0.7--7 keV. However, given the
large uncertainty of the above estimate and the possible contribution
of star-forming activity to the narrow emission lines, we cannot
completely rule out that a persistent weak hard X-ray component is
present but is too weak to be detected. Alternatively, it could be
possible that the TDE has destroyed the corona or jet that is
responsible for hard X-ray emission.

\section*{Acknowledgments}

We thank the anonymous referee for the helpful
comments. DL is supported by the National Aeronautics and Space Administration
through Chandra Award Number GO5-16114X issued by the Chandra X-ray
Observatory Center, which is operated by the Smithsonian Astrophysical
Observatory for and on behalf of the National Aeronautics Space
Administration under contract NAS8-03060. The work of LCH was
supported by National Key Program for Science and Technology Research
and Development grant 2016YFA0400702. We thank the Swift PI Neil
Gehrels for approving our ToO request to make several observations of
2XMM J123103.2+110648.

Funding for SDSS-III has been provided by the Alfred P. Sloan Foundation, the Participating Institutions, the National Science Foundation, and the U.S. Department of Energy Office of Science. The SDSS-III web site is http://www.sdss3.org/. SDSS-III is managed by the Astrophysical Research Consortium for the Participating Institutions of the SDSS-III Collaboration including the University of Arizona, the Brazilian Participation Group, Brookhaven National Laboratory, Carnegie Mellon University, University of Florida, the French Participation Group, the German Participation Group, Harvard University, the Instituto de Astrofisica de Canarias, the Michigan State/Notre Dame/JINA Participation Group, Johns Hopkins University, Lawrence Berkeley National Laboratory, Max Planck Institute for Astrophysics, Max Planck Institute for Extraterrestrial Physics, New Mexico State University, New York University, Ohio State University, Pennsylvania State University, University of Portsmouth, Princeton University, the Spanish Participation Group, University of Tokyo, University of Utah, Vanderbilt University, University of Virginia, University of Washington, and Yale University.

\section*{REFERENCES}
\bibliographystyle{mn2e}  
%\bibliography{all-bib}

\bsp

\label{lastpage}

\end{document}